\documentclass[conference]{IEEEtran}
\usepackage{amsmath}
\usepackage{latexsym}
\usepackage{graphicx}
\usepackage{bbding}
\usepackage{color}
\IEEEoverridecommandlockouts
\begin{document}

\title{Adaptive Mode Selection in Multiuser MISO Cognitive Networks with Limited Cooperation and Feedback}

\author{\normalsize
Xiaoming~Chen, Zhaoyang~Zhang and Chau~Yuen
\thanks{Copyright (c) 2013 IEEE. Personal use of this material is
permitted. However, permission to use this material for any other
purposes must be obtained from the IEEE by sending a request to
pubs-permissions@ieee.org.}
\thanks{This work was supported by the National Nature Science
Foundation (No. 61301102), the Natural Science Foundation of Jiangsu
Province (No. BK20130820), the open research fund of State Key
Laboratory of Integrated Services Networks, Xidian University (No.
ISN14-01) and the Doctoral Fund of Ministry of Education of China
(No. 20123218120022).}
\thanks{Xiaoming~Chen (e-mail: {\tt chenxiaoming@nuaa.edu.cn})
is with the College of Electronic and Information Engineering,
Nanjing University of Aeronautics and Astronautics, and also with
the State Key Laboratory of Integrated Services Networks, Xidian
University, China. Zhaoyang~Zhang (e-mail: {\tt
ning\_ming@zju.edu.cn}) is with the Department of Information
Science and Electronic Engineering, Zhejiang University, China.
Chau~Yuen ({\tt yuenchau@sutd.edu.sg}) is with the Singapore
University of Technology and Design, Singapore.}} \maketitle

\begin{abstract}
In this paper, we consider a multiuser MISO downlink cognitive
network coexisting with a primary network. With the purpose of
exploiting the spatial degree of freedom to counteract the
inter-network interference and intra-network (inter-user)
interference simultaneously, we propose to perform zero-forcing
beamforming (ZFBF) at the multi-antenna cognitive base station (BS)
based on the instantaneous channel state information (CSI). The
challenge of designing ZFBF in cognitive networks lies in how to
obtain the interference CSI. To solve it, we introduce a limited
inter-network cooperation protocol, namely the quantized CSI
conveyance from the primary receiver to the cognitive BS via
purchase. Clearly, the more the feedback amount, the better the
performance, but the higher the feedback cost. In order to achieve a
balance between the performance and feedback cost, we take the
maximization of feedback utility function, defined as the difference
of average sum rate and feedback cost while satisfying the
interference constraint, as the optimization objective, and derive
the transmission mode and feedback amount joint optimization scheme.
Moreover, we quantitatively investigate the impact of CSI feedback
delay and obtain the corresponding optimization scheme. Furthermore,
through asymptotic analysis, we present some simple schemes.
Finally, numerical results confirm the effectiveness of our
theoretical claims.
\end{abstract}

\begin{keywords}
Cognitive network, limited cooperation, feedback utility, feedback
delay, mode selection.
\end{keywords}

\section{Introduction}
One challenging problem that hinders the further development of
wireless communications is the scarcity of available spectrum for
the demands of various advanced wireless services. However, FCC's
report shows that the licensed spectrum is heavily underutilized in
temporal, frequency or spatial scales \cite{FCC}. A good solution to
solve the above contradiction is the use of cognitive radio (CR)
\cite{Mitola} \cite{Haykin}. Generally speaking, a cognitive network
is allowed to coexist with a primary network on the licensed
spectrum by adaptively adjusting its transmit parameters, i.e.,
power, time, and frequency, so that the resulting interference to
the primary network meets a certain constraint.

Intuitively, the design objective of a CR network is to maximize the
spectrum efficiency while satisfying the given interference
constraint. Considering the dynamic activity of the primary network,
one feasible method is to access the idle spectrum in temporal or
frequency scale through effective spectrum sensing
\cite{sensing1}-$\!\!$\cite{sensing3}. Clearly, the performance of
this method is dependent on the precision of spectrum sensing to a
large extent. In fact, it is a passive way for spectrum efficiency
improvement, because the spectrum is available only when primary
user is inactive. A better method is to find the spectrum
availability actively by making use of channel fading, no matter
whether the spectrum is idle. As a simple example, the author in
\cite{PA} proposed to opportunistically control transmit power
according to the instantaneous channel state information (CSI), so
as to optimize the spectrum efficiency. The key of the realization
of active spectrum share lies in exploiting the extra degrees of
freedom other than time and frequency. Inspired by interference
avoidance through making use of spatial degree of freedom
\cite{Multiantenna1} \cite{Multiantenna2}, the combination of CR and
multiantenna technologies has been proved as an advanced way to
achieve the goal of CR in previous literatures
\cite{CogMIMO1}-$\!\!$\cite{CogMIMO3}. In this paper, we focus on
the design of a multi-antenna transmission scheme for the multiuser
MISO cognitive networks under limited CSI feedback.

\subsection{Related Works}
The common to the previous works on multi-antenna cognitive networks
lies in how to achieve a balance between increasing the transmission
rate of cognitive network and decreasing the interference to primary
network. A simplest method to tackle such a challenge is to transmit
the signal of cognitive network in the null space of the
interference channel from cognitive transmitter to primary receiver.
It makes sense only when primary network can not bear any
interference. If the interference is bearable within a given
constraint, it has been proved that the optimal transmit direction,
namely transmit beam, lies in the space spanned jointly by
interference channel and the projection of cognitive channel from
cognitive transmitter to cognitive receiver into interference
channel \cite{MA1}. Considering the availability of interference CSI
may be imperfect, an adaptive beamforming design method coupling
with transmission mode selection was proposed, so that the QoS
requirement can be fulfilled with the minimum resource consumption
\cite{MA2}. In addition, a robust beamforming method based on mean
and covariance information of interference channel was given in
\cite{MA3}. With the purpose of further improving spectrum
efficiency, the research focus is shifting from single user to
multiuser. The authors in \cite{Multiuser1} first chose the users
whose channels are quasi-orthogonal to interference channel so that
the interference are controlled, then a given number of users whose
channels are as orthogonal as possible are selected from the above
user set, so as to decrease the inter-user interference in cognitive
networks. Similarly, an user scheduling scheme was presented to
improve the performance of cognitive networks by exploiting the
so-called multiuser interference diversity \cite{Multiuser2}.
Furthermore, a joint power control and beamforming strategy was
proposed to minimize the total transmit power while satisfying
interference constraint to primary users and quality-of-service
(QoS) requirements of cognitive users simultaneously
\cite{Multiuser3}.

All the above multiantenna schemes can improve spectrum efficiency
while meeting interference constraint to some extent. However, they
all fend off the question that how multiantenna cognitive
transmitter knows the interference CSI. Naturally, the interference
CSI is critical to beamforming design and user scheduling at the
transmitter. In traditional multiantenna systems, the CSI is usually
obtained by limited feedback from the receiver to the transmitter
through a quantization codebook \cite{Feedback1} \cite{Feedback2}.
Yet, in cognitive networks, the achievement of the interference CSI
needs the cooperation from primary receiver. Inspired by this idea,
cooperative feedback from primary receiver to cognitive transmitter
was first proposed in \cite{CooperativeFeedback1}, so the cognitive
transmitter can design the optimal beamformer according to the
instantaneous interference CSI. Meanwhile, the authors also gave a
detailed discussion on feedback allocation between channel direction
information and interference power control information, so as to
optimize the performance while meeting the interference constraint.
Similarly, based on the cooperative feedback information, the
authors in \cite{CooperativeFeedback2} proposed to select the
optimal cognitive user, such that the benefit of multiuser diversity
can be obtained. Intuitively, the more accurate the CSI at the
transmitter, the better the system performance. However, different
from traditional multiantenna systems, primary receiver has no
obligation to convey the CSI to cognitive transmitter. This is an
inevitable problem in cooperative feedback based cognitive networks.

Moreover, transmission mode, namely the number of accessing users,
is also an important factor for performance improvement in multiuser
multiantenna system \cite{ModeSelection1} \cite{ModeSelection2}. To
be precise, a large transmission mode can obtain high spatial
multiplexing gain, while it may result in severe inter-user
interference. Hence, it is necessary to select the optimal
transmission mode according to channel conditions, especially in the
case of limited CSI feedback, since the amount of CSI has a great
impact on mode selection. In \cite{Modeselection3}, the authors gave
a comprehensive performance analysis for a multiuser MISO downlink
with limited feedback and feedback delay, then proposed mode
switching between single-user (SU) mode and multiuser (MU) mode by
comparing the average sum rates for a given SINR. It is worth
pointing out this work only considers two transmission mode, name
$M=1$ and $M=N_t$, where $N_t$ is the number of transmit antenna at
BS. In multiuser multiantenna cognitive network, mode selection is
an effective way of coordinating intra-network (inter-user) and
inter-network interference, and thus improve the performance
significantly. Yet, mode selection in cognitive network is a more
challenging task, since it is affected by intra-network interference
and inter-network interference simultaneously, especially when the
CSI is imperfect.

\subsection{Main Contributions}
In this paper, we focus on a limited cooperative multiuser MISO
cognitive network. In order to let primary receiver help cognitive
transmitter of its own accord, we propose that cognitive network
purchases partial interference CSI from primary network with an
appreciate cost. Therefore, primary network obtains some profits and
cognitive network improves its performance, which is a win-win
solution. Intuitively, the more the feedback amount, the better the
performance, but the higher the feedback cost. For the sake of
balancing system performance and feedback cost, it is imperative to
design a feedback efficient multiuser transmission scheme. Thereby,
we propose a concept of feedback utility function, defined as the
difference of the average sum rate of cognitive network and feedback
cost. In this paper, we take the maximization of feedback utility
function as the optimization objective. As analyzed above, for a
multiantenna broadcast channel with limited feedback, the number of
accessing users, namely transmission mode, has a great impact on the
average sum rate. Hence, in order to maximize the feedback utility
function, it is necessary to jointly optimize feedback amount and
transmission mode. The main contributions of this paper are
summarized as follows
\begin{enumerate}
\item We propose a framework of limited cooperative zero-forcing
beamforming (ZFBF) for multiuser MISO cognitive networks, so as to
mitigate the inter-network and intra-network interferences
effectively.

\item We reveal the relationship among the interference, feedback
amount, transmit power and number of accessing user, so that the
parameters of cognitive network can be adaptively adjusted according
to interference constraint.

\item We define a feedback utility function as the difference of the average sum
rate of cognitive network and feedback cost. By maximizing the
feedback utility function, we get the optimal and suboptimal
algorithms that derive the transmission mode and feedback amount.

\item We investigate of the impact of feedback delay on
feedback utility function, and obtain analytically the corresponding
transmission mode and feedback amount joint optimization scheme.

\item We present some relatively simple optimization schemes in three extreme
cases through asymptotic performance analysis.

\end{enumerate}

\subsection{Paper Organization}
The rest of this paper is organized as follows. In Section II, we
give a brief overview of the system model and introduce the
transmission protocol of cognitive network. Section III focuses on
the derivation of the optimal feedback amount and transmission mode
by maximizing the feedback utility function while satisfying the
interference constraint. Then, the asymptotic characteristics of the
performance is analyzed in Section IV. Simulation results are given
in Section V to validate the effectiveness of the proposed schemes
and we conclude the whole paper in Section VI.

\textit{Notation}: We use bold upper (lower) letters to denote
matrices (column vectors), $(\cdot)^H$ to denote conjugate
transpose, $E[\cdot]$ to denote expectation, $\|\cdot\|$ to denote
the $L_2$ norm of a vector, $|\cdot|$ to denote the absolute value,
$\lceil a\rceil$ to denote the smallest integer not less than $a$,
$\lfloor a\rfloor$ to denote the largest integer not greater than
$a$, and $\stackrel{d}{=}$ to denote the equality in distribution.
The acronym i.i.d means ``independent and identically distributed",
pdf means ``probability density function" and cdf means ``cumulative
density function".

\section{System Model}
\begin{figure}[h] \centering
\includegraphics [width=0.45\textwidth] {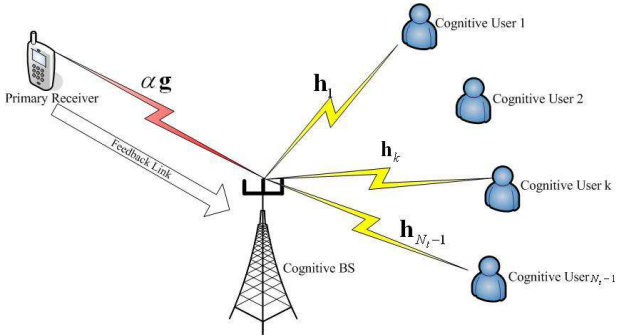}
\caption {The considered system model.} \label{Fig1}
\end{figure}

We consider a system composed of a primary network and a cognitive
network, as seen in Fig.\ref{Fig1}. It is assumed that the two
networks share the same spectrum for transmission. The cognitive
network includes an $N_t$-antenna cognitive base station (BS) and
$K$ single-antenna cognitive users. In order to guarantee the normal
communication of primary network, the interference from cognitive BS
to the single-antenna primary receiver must satisfy a given
constraint, which is the precondition that primary network allows
cognitive network to access the licensed spectrum. Meanwhile,
primary transmitter will interfere with cognitive network. In this
paper, we integrate the interference from primary transmitter into
the noise at cognitive users, so we omit primary transmitter in this
system. Note that this paper focuses on mode selection but not user
scheduling, we fix the user number $K=N_t-1$ because at most $N_t-1$
users are allowed to access cognitive network while one spatial
degree of freedom of $N_t$-antennas cognitive BS is used for
interference suppression. In practical network, if $K$ is large, we
can combine space division multiplexing access (SDMA) with time
division multiplexing access (TDMA). Specifically, cognitive users
can access the network in different time slots based on some user
scheduling schemes.

It is assumed that the downlink is homogenous, where cognitive users
have the same channel vector statistics, same path loss and same
receive noise variance. The channels $\textbf{h}_k, k=1,\cdots,
N_t-1$ from cognitive BS to cognitive users are $N_t$ dimensional
i.i.d Gaussian random vectors with zero mean and unit variance, and
the path loss is normalized to 1. The path loss component of
interference channel from cognitive BS to primary receiver is
$\alpha$ with respect to that of downlink channel, and small scale
fading component $\textbf{g}$ is also an $N_t$ dimensional circular
symmetrical complex Gaussian random vector. Considering path loss
varies slowly and is a scalar, we assume cognitive BS knows $\alpha$
fully through the aid of a positioning system and the amount of
exchange information can be neglected. Cognitive network
communicates in the form of time slot, and all the channels are
assumed to keep constant during one time slot and fade independently
slot by slot. At the beginning of each time slot, primary receiver
and cognitive users convey the corresponding CSI to cognitive BS
based on quantization codebooks. All the codebooks are designed in
advance and stored at the transmitter and the receivers. Assuming
the codebook of size $2^B$ at primary receiver is
$\mathcal{G}=\{\hat{\textbf{g}}_1,\cdots,\hat{\textbf{g}}_{2^B}\}$,
which is a collection of unit norm vectors, then the optimal
quantization codeword selection criterion can be expressed as
\begin{equation}
i=\arg\max\limits_{1\leq
j\leq2^B}|\hat{\textbf{g}}_j^{H}\tilde{\textbf{g}}|^2,\label{eqn1}
\end{equation}
where $\tilde{\textbf{g}}=\frac{\textbf{g}}{\|\textbf{g}\|}$ is the
channel direction vector. Specifically, index $i$ is conveyed by
primary receiver and $\hat{\textbf{g}}_i$ is recovered at cognitive
BS as instantaneous CSI of the interference channel. The
quantization codebook of size $2^C$ at cognitive user $k$ is
$\mathcal{H}_k=\{\hat{\textbf{h}}_{k,1},\cdots,\hat{\textbf{h}}_{k,2^C}\},k=1,\cdots,N_t-1$,
and the same process of codeword selection, index conveyance and CSI
recovery is used to help cognitive BS to obtain CSI of downlink
channels. It is worth pointing out that the feedback amount $C$ is
fixed and $B$ is dynamically variable as discussed below.

Based on the feedback information from primary receiver and
cognitive users, cognitive BS determines the optimal transmit beam
of the $m$th user of the selected $M$ users
$\textbf{w}_{K_m},m=1,\cdots,M,1\leq M\leq N_t-1$ by making use of
zero-forcing beamforming design methods, when given transmission
mode $M$. Specifically, for $\hat{\textbf{h}}_{K_m}$, the
complementary channel matrix is
\begin{equation}
\bar{\hat{\textbf{H}}}_{K_m}=[\hat{\textbf{g}},\hat{\textbf{h}}_{K_1},\cdots,\hat{\textbf{h}}_{K_{m-1}},
\hat{\textbf{h}}_{K_{m+1}},\cdots,\hat{\textbf{h}}_{K_M}].\nonumber
\end{equation}
Taking singular value decomposition (SVD) to
$\bar{\hat{\textbf{H}}}_{K_m}$, if $\textbf{V}_{K_m}^{\perp}$ is the
matrix composed of the right singular vectors related to zero
singular values, then we randomly choose a unit norm vector from the
space spanned by $\textbf{V}_{K_m}^{\perp}$ as the transmit beam
$\textbf{w}_{K_m}$. Since $\textbf{V}_{K_m}^{\perp}$ is the null
space of $\bar{\hat{\textbf{H}}}_{K_m}$, we have
\begin{equation}
\hat{\textbf{g}}^{H}\textbf{w}_{K_m}=0,\label{eqn2}
\end{equation}
and
\begin{equation}
\hat{\textbf{h}}_{K_k}^{H}\textbf{w}_{K_m}=0,\label{eqn3}
\end{equation}
where $1\leq k \leq M, k\neq m$. Thus, the receive signal at the
$K_m$th user can be expressed as
\begin{equation}
y_{K_m}=\sqrt{\frac{P}{M}}\textbf{h}_{K_m}^H\sum\limits_{k=1}^{M}\textbf{w}_{K_k}x_{K_k}+n,\label{eqn28}
\end{equation}
where $x_{K_m}$ the normalized transmit signal related to the
$K_k$th user, $y_{K_m}$ is the receive signal at the $K_m$th user,
$P$ is the total transmit power of cognitive BS, which is
distributed to $M$ users equally, and $n$ is additive white Gaussian
noise with zero mean and variance $\sigma^2$. Then, the interference
from cognitive BS to primary receiver, and the receive signal to
interference and noise ratio (SINR) at cognitive user $K_m$ are
given by
\begin{equation}
I=P/M\alpha\sum\limits_{m=1}^{M}|\textbf{g}^H\textbf{w}_{K_m}|^2,\label{eqn4}
\end{equation}
and
\begin{equation}
SINR_{K_m}=\frac{|\textbf{h}_{K_m}^H\textbf{w}_{K_m}|^2}{\sigma^2M/P+\sum\limits_{k=1,k\neq
m}^{M}|\textbf{h}_{K_m}^H\textbf{w}_{K_k}|^2},\label{eqn5}
\end{equation}
respectively.

The precondition that cognitive network is allowed to access the
licensed spectrum is that the interference from cognitive BS to
primary receiver meets a given constraint. Average interference
constraint (AIC) and peak interference constraint (PIC) are two
commonly used interference conditions in cognitive networks.
Previous works have proved that from the view of throughput
maximization, given the same average and peak interference
constraints, the AIC is more favorable than the PIC because of its
more flexibility for dynamically allocating transmit powers over
different fading states \cite{AIC1} \cite{AIC2}, thus we take the
AIC as the constraint condition in this paper. If
$\bar{I}_{\textrm{AIC}}$ is the allowable maximum average
interference, the interference constraint is given by
\begin{equation}
\bar{I}=E[I]\leq\bar{I}_{\textrm{AIC}}.\label{eqn29}
\end{equation}

Intuitively, when $\bar{I}_{\textrm{AIC}}$ is given, the available
maximum transmit power $P$ increases with the increase of feedback
amount $B$, and thus the performance of cognitive network improves
accordingly, but the feedback cost also increases. In order to
achieve a tradeoff between the performance and feedback cost, we
take the maximization of feedback utility function as the
optimization objective, which is defined as
\begin{eqnarray}
&&f(M,B)\nonumber\\&=&E\Bigg[\sum\limits_{m=1}^{M}\log_2\left(1+\frac{|\textbf{h}_{K_m}^H\textbf{w}_{K_m}|^2}{\sigma^2M/P+\sum\limits_{k=1,k\neq
m}^{M}|\textbf{h}_{K_m}^H\textbf{w}_{K_k}|^2}\right)\nonumber\\&-&\mu
B\Bigg]\nonumber
\end{eqnarray}
\begin{eqnarray}
&=&ME\left[\log_2\left(1+\frac{|\textbf{h}_{K_1}^H\textbf{w}_{K_1}|^2}{\sigma^2M/P+\sum\limits_{k=2}^{M}|\textbf{h}_{K_1}^H\textbf{w}_{K_k}|^2}\right)\right]-\mu
B,\nonumber\\\label{eqn9}
\end{eqnarray}
where (\ref{eqn9}) holds true since the $SINR$s of the users are
i.i.d. $\mu$ is the pricing factor of feedback information from the
primary receiver whose unit is rate penalty per feedback bit. $\mu$
is depending on cooperative cost, mainly including power and
spectrum consumption for channel estimation, codeword selection and
index feedback, and is determined based on a negotiation or
competition scheme between cognitive and primary networks, such as
stackelberg game \cite{Stackelberg}. Generally speaking, if primary
network asks for a higher unit price, cognitive network will
purchase less feedback information, resulting in low profits. The
optimal $\mu$ will be determined when one side can not further
improve its utility, if the other side's utility is not deceased.
Once the price is determined, the owner of cognitive network will
pay the owner of primary network. Thus, these two networks can get
some profits from cooperative feedback. In this paper, we
concentrate on the optimization of feedback utility, so we assume
the pricing factor has been determined by the two sides based on a
certain method, such as stackelberg game. As seen in (\ref{eqn9}),
feedback utility function is a function of $B$ and $M$. In order to
maximize feedback utility function, it is necessary to jointly
optimize the two variables. Hence, the focus of this paper is on the
joint optimization of feedback amount $B$ and transmission mode $M$
by maximizing feedback utility function (\ref{eqn9}) while
satisfying AIC.

\section{Adaptive Mode Selection}
As mentioned earlier, this paper focuses on the maximization of the
feedback utility function while satisfying AIC through adaptive mode
selection and dynamic feedback control. As seen in (\ref{eqn4}),
given a channel condition $\alpha$, the interference $I$ is a
function of transmit power $P$, transmission mode $M$ and feedback
amount $B$. Hence, when $I$ is given, there is a tradeoff among $P$,
$M$ and $B$. On the other hand, it is known from (\ref{eqn5}) that,
$P$ and $M$ also have a great impact on the receive SINR, and thus
the feedback utility function. Therefore, prior to discussing the
optimal transmission mode $M$, we need to reveal the relationship
among the interference, feedback utility function, transmit power,
feedback amount and transmission mode.

By using ZFBF based on limited feedback information from primary
receiver, the relation between the residual AIC, transmit power,
transmission mode and feedback amount is given by

\emph{Theorem 1}: Given $P$ and $B$, the residual AIC at primary
receiver after ZFBF is tightly upper bounded by $\frac{\alpha
PN_t}{N_t-1}2^{-\frac{B}{N_t-1}}$.

\begin{proof}
Following the theory of random vector quantization (RVQ) \cite{RVQ},
the relationship between the original and the quantized channel
direction vectors can be expressed as
\begin{equation}
\tilde{\textbf{g}}=\sqrt{1-a}\hat{\textbf{g}}+\sqrt{a}\textbf{s},\label{eqn6}
\end{equation}
where $\hat{\textbf{g}}$ is the optimal quantization codeword based
on codeword selection criterion (\ref{eqn1}),
$a=\sin^2\left(\angle\left(\tilde{\textbf{g}},\hat{\textbf{g}}\right)\right)$
is the magnitude of the quantization error, and $\textbf{s}$ is an
unit norm vector isotropically distributed in the nullspace of
$\hat{\textbf{g}}$, and is independent of $a$. Therefore, the $m$th
term of the sum in the right hand side (RHS) of (\ref{eqn4}) is
derived as
\begin{eqnarray}
|\tilde{\textbf{g}}^H\textbf{w}_{K_m}|^2&=&\left|\sqrt{1-a}\hat{\textbf{g}}^H\textbf{w}_{K_m}+\sqrt{a}\textbf{s}^H\textbf{w}_{K_m}\right|^2\nonumber\\
&=&a|\textbf{s}^H\textbf{w}_{K_m}|^2,\label{eqn7}
\end{eqnarray}
where (\ref{eqn7}) follows from the fact that $\textbf{w}_{K_m}$ is
in the nullspace of $\hat{\textbf{g}}$, namely
$\hat{\textbf{g}}^H\textbf{w}_{K_m}=0$. Due to the independence of
channel magnitude component and channel direction component, the AIC
is reduced to
\begin{eqnarray}
\bar{I}&=&\alpha
P/M\sum\limits_{m=1}^ME[\|\textbf{g}\|^2]E[a]E[|\textbf{s}^H\textbf{w}_{K_m}|^2]\nonumber\\
&=&\alpha
PE[\|\textbf{g}\|^2]E[a]E[|\textbf{s}^H\textbf{w}_{K_1}|^2],\label{eqn8}
\end{eqnarray}
where (\ref{eqn8}) holds true because
$|\textbf{s}^H\textbf{w}_{K_m}|^2$ is i.i.d. for $m=1,\cdots,M$.
Hence, in order to obtain $\bar{I}$, we only need to compute the
three expectation terms. First, since $\|\textbf{g}\|^2$ is
$\chi^2(2N_t)$ distributed, we have $E[\|\textbf{g}\|^2]=N_t$. As we
known, $a=1-|\hat{\textbf{g}}^H\tilde{\textbf{g}}|^2$. For an
arbitrary quantization codeword $\hat{\textbf{g}}_j$,
$1-|\hat{\textbf{g}}_j^H\tilde{\textbf{g}}|^2$ is $\beta(N_t-1,1)$
distributed \cite{RVQ}, so that $a$ is the minimum of $2^B$
independent $\beta(N_t-1,1)$ random variable, and thus its
expectation can be tightly upper bounded as
$E[a]<2^{-\frac{B}{N_t-1}}$. Finally, since $\textbf{s}$ and
$\textbf{w}_{K_1}$ are i.i.d. isotropic vectors in the $(N_t-1)$
dimensional nullspace of $\hat{\textbf{g}}$,
$|\textbf{s}^H\textbf{w}_{K_1}|^2$ is $\beta(1, N_t-2)$ distributed
\cite{RVQ}, whose expectation is equal to $\frac{1}{N_t-1}$. As a
result, we have $\bar{I}<\frac{\alpha
PN_t}{N_t-1}2^{-\frac{B}{N_t-1}}$
\end{proof}

According to Theorem 1, given AIC $I_{\textrm{AIC}}$, feedback
amount $B$ and transmit mode $M$, in order to meet AIC strictly,
transmit power $P$ has an upper bound
\begin{equation}
P=\frac{(N_t-1)I_{\textrm{AIC}}}{\alpha
N_t}2^{\frac{B}{N_t-1}}.\label{eqn10}
\end{equation}

Substituting (\ref{eqn10}) into (\ref{eqn9}), we have
\begin{eqnarray}
&&f(M,B)\nonumber\\&=&ME\left[\log_2\left(1+\frac{|\textbf{h}_{K_1}^H\textbf{w}_{K_1}|^2}{\eta
M2^{-\frac{B}{N_t-1}}+\sum\limits_{k=2}^{M}|\textbf{h}_{K_1}^H\textbf{w}_{K_k}|^2}\right)\right]\nonumber\\
&-&\mu B,\label{eqn11}
\end{eqnarray}
where $\eta=\frac{\alpha N_t\sigma^2}{(N_t-1)I_{\textrm{AIC}}}$.
Clearly, the key of obtaining $f(M,B)$ is to compute the average
rate of the $K_m$th cognitive user. Before computing the average
rate, we need to know the pdf of the receive $SINR$. Examining the
receive $SINR_{K_1}$, it is upper bounded by

\begin{eqnarray}
SINR_{K_1}&=&\frac{|\textbf{h}_{K_1}^H\textbf{w}_{K_1}|^2}{\eta
M2^{-\frac{B}{N_t-1}}+\sum\limits_{k=2}^{M}|\textbf{h}_{K_1}^H\textbf{w}_{K_k}|^2}\nonumber\\
&=&\frac{\|\textbf{h}_{K_1}\|^2|\tilde{\textbf{h}}_{K_1}^H\textbf{w}_{K_1}|^2}{\eta
M2^{-\frac{B}{N_t-1}}+\|\textbf{h}_{K_1}\|^2b\sum\limits_{k=2}^{M}|\textbf{s}^H\textbf{w}_{K_k}|^2}\label{eqn12}\\
&\stackrel{d}{=}&\frac{\|\textbf{h}_{K_1}\|^2\beta(1,N_t-1)}{\eta
M2^{-\frac{B}{N_t-1}}+\|\textbf{h}_{K_1}\|^2b\sum\limits_{k=2}^{M}\beta(1,N_t-2)}\label{eqn13}\\
&\stackrel{d}{=}&\frac{\|\textbf{h}_{K_1}\|^2\beta(1,N_t-1)}{\eta
M2^{-\frac{B}{N_t-1}}+\sum\limits_{k=2}^{M}\Gamma(N_t-1,\delta)\beta(1,N_t-2)}\label{eqn14}\\
&\stackrel{d}{=}&\frac{\chi_2^2}{\eta
M2^{-\frac{B}{N_t-1}}+\delta\chi_{2(M-1)}^2}\nonumber
\end{eqnarray}
where
$b=\sin^2(\angle(\tilde{\textbf{h}}_{K_1},\hat{\textbf{h}}_{K_1}))$
and $\delta=2^{-\frac{C}{N_t-1}}$. (\ref{eqn12}) holds true because
$\tilde{\textbf{h}}_{K_1}=\sqrt{1-b}\hat{\textbf{h}}_{K_1}+\sqrt{b}\textbf{s}$
and $\hat{\textbf{h}}_{K_1}^H\textbf{w}_{K_k}=0$, (\ref{eqn13})
follows from the fact that
$|\tilde{\textbf{h}}_{K_1}^H\textbf{w}_{K_1}|^2$ and
$|\textbf{s}^H\textbf{w}_{K_k}|^2$ are $\beta(1,N_t-1)$ and
$\beta(1,N_t-2)$ distributed as analyzed in (\ref{eqn8}), and
(\ref{eqn14}) is derived since $\|\textbf{h}_{K_1}\|^2b$ is
$\Gamma(N_t-1,\delta)$ distributed according to the theory of
quantization cell approximation \cite{QCA}. For the product of a
$\Gamma(N_t-1,\delta)$ distributed random variable and a
$\beta(1,N_t-2)$ distributed random variable, it is equal to
$\delta\chi_2^2$ in distribution \cite{Modeselection3}. Based on the
fact that the sum of $M-1$ independent $\chi_2^2$ distributed random
variables is $\chi_{2(M-1)}^2$ distributed, $SINR_{K_1}$ is equal to
$\frac{\chi_2^2}{\eta M2^{-\frac{B}{N_t-1}}+\delta\chi_{2(M-1)}^2}$
in distribution. Let $y\sim\chi_{2(N_t-1)}^2$ and $z\sim\chi_2^2$,
we can derive the cdf and pdf of $SINR_{K_m}$ as
\begin{eqnarray}
P(x)&=&P\left(\frac{z}{\eta
M2^{-\frac{B}{N_t-1}}+\delta y}\leq x\right)\nonumber\\
&=&\int_0^{\infty}F_{Z|Y}\big(x(\eta
M2^{-\frac{B}{N_t-1}}+\delta y)\big)f_Y(y)dy\nonumber\\
&=&\int_0^{\infty}\big(1-\exp(-x(\eta M2^{-\frac{B}{N_t-1}}+\delta
y))\big)\nonumber\\
&\times&\frac{y^{M-2}}{\Gamma(M-1)}\exp(-y)dy\nonumber\\
&=&1-\frac{\exp\left(-\eta
M2^{-\frac{B}{N_t-1}}x\right)}{\left(1+\delta
x\right)^{M-1}}\nonumber\\
&\times&\int_0^{\infty}\frac{\exp(-(1+\delta x)y)((1+\delta x)y)^{M-2}}{\Gamma(M-1)}d((1+\delta x)y)\nonumber\\
&=&1-\frac{\exp\left(-\eta
M2^{-\frac{B}{N_t-1}}x\right)}{\left(1+\delta
x\right)^{(M-1)}},\label{eqn27}
\end{eqnarray}
and
\begin{eqnarray}
p(x)&=&\eta M2^{-\frac{B}{N_t-1}}\exp\left(-\eta
M2^{-\frac{B}{N_t-1}}x\right)\left(1+\delta
x\right)^{-(M-1)}\nonumber\\
&-&\delta(N_t-1)\exp\left(-\eta
M2^{-\frac{B}{N_t-1}}x\right)\left(1+\delta
x\right)^{-M},\label{eqn15}
\end{eqnarray}
respectively, where $F_{Z|Y}(\cdot)$ is the conditional cdf of $z$
for a given $y$, $f_Y(\cdot)$ is the pdf of $y$, $\Gamma(\cdot)$ is
the Gamma function. (\ref{eqn27}) holds true because
$\frac{\exp(-(1+\delta x)y)((1+\delta x)y)^{M-2}}{\Gamma(M-1)}$ is
the pdf of $(1+\delta x)y$. Hence, $f(M,B)$ can be casted as
\begin{eqnarray}
&&f(M,B)\nonumber\\&=&M\int_0^\infty\log_2\left(1+x\right)p(x)dx-\mu
B\nonumber\\
&\approx&M\log2(e)E\left[\ln(1+x)\right]-\mu
B\nonumber\\
&=&M\log2(e)\int_{0}^{\infty}\frac{1-P(x)}{x+1}dx-\mu
B\nonumber\\
&=&\frac{\log_2(e)M}{\delta^{M-1}}\int_{0}^{\infty}\frac{\exp\left(-\eta
M2^{-\frac{B}{N_t-1}}x\right)}{(x+1)(x+\delta^{-1})^{M-1}}dx-\mu
B\nonumber\\
&=&\frac{\log_2(e)M}{\delta^{M-1}}I_1\left(\eta
M2^{-\frac{B}{N_t-1}}, \delta^{-1}, M-1\right)-\mu B,\label{eqn16}
\end{eqnarray}
where
\begin{eqnarray}
I_1(x,y,z)&=&\int_{0}^{\infty}\frac{\exp\left(-xt\right)}{(t+1)(t+y)^z}dt\nonumber\\
&=&\sum\limits_{i=1}^{z}(-1)^{i-1}(1-y)^{-i}I_2(x,y,
z-i+1)\nonumber\\&+&(y-1)^{-z}I_2(x,1,1),\nonumber
\end{eqnarray}
and
\begin{equation}
I_2(x,y,z)=\left\{\begin{array}{ll} \exp(xy)\textmd{E}_1(xy) & z=1\\
\sum\limits_{k=1}^{z-1}\frac{(k-1)!}{(z-1)!}\frac{(-x)^{z-k-1}}{y^k}\\+\frac{(-x)^{z-1}}{(z-1)!}\exp(xy)E_1(xy)
& z\geq2
\end{array}\right.\nonumber
\end{equation}
and $\textmd{E}_1(x)$ is the exponential-integral function of the
first order.

So far, we have derived the closed-form expression of feedback
utility function $f(M,B)$ as a function of $M$ and $B$. Then, the
joint optimization of $M$ and $B$ is equivalent to the following
optimization problem:
\begin{eqnarray}
J_1: \max\limits_{M,B} &&f(M,B)\nonumber\\
s.t.\quad &&2\leq M\leq N_t-1\nonumber\\
&&B\leq B_0,\nonumber
\end{eqnarray}
where $B_0$ is the maximum usable feedback amount considering the
complexity of codebook design and codeword selection in practical
networks. $M$ must be greater than $1$ because $M=1$ is unsuitable
for ZFBF. Since both $M$ and $B$ are integers, $J_1$ is an integer
programming problem, so it is difficult to get the closed-form
expressions of $M$ and $B$. Considering $M$ and $B$ are upper
bounded by $N_t-1$ and $B_0$ respectively, which are impossibly to
be large values in practical networks. Thereby, for each
transmission mode, we could derive the optimal feedback bits and the
corresponding feedback utility function. Among all $(M,B)$
combinations, it is easy to get the optimal one with the maximum
feedback utility function. Thus, the whole procedure can be
summarized as below.\\

\rule{8.45cm}{1pt}

\emph{Optimal Algorithm}
\begin{enumerate}
\item Initialization: given $N_t$, $\mu$, $\alpha$, $B_0$, $C$ and
$I_{\textrm{AIC}}$. Set $M=2$, $B=0$, $i=1$ and $j=1$.

\item Let $M_i=M$, $B_j=B$, and compute $f_{i,j}=f(M,B)$.

\item If $B+1\leq B_0$, then $B=B+1$, $j=j+1$, and go to 2).

\item If $M+1\leq N_t-1$, then $M=M+1$, $i=i+1$, $B=0$, $j=1$ and go to 2).

\item Search $(n^{*},m^{*})=\arg\min\limits_{1\leq n\leq i,1\leq m\leq j}f_{n,m}$.
$M_{n^{*}}$ and $B_{m^{*}}$ are the optimal transmission mode and
feedback amount, respectively.

\end{enumerate}

\rule{8.45cm}{1pt}

Clearly, the optimal algorithm essentially is a numerical search
method with high complexity. Herein, we give a suboptimal algorithm
with low complexity to derive the $(M,B)$ combination. As mentioned
earlier, one of the difficulties to solve the above problem lies in
that both $M$ and $B$ are integers. If we relax the two constraints,
we may reduce the solving complexity. In this context, $J_1$ is
transformed as the following general optimization problem
\begin{eqnarray}
J_2: \max\limits_{M,B} &&f(M,B)\nonumber\\
s.t.\quad &&2\leq M\leq N_t-1\nonumber\\
&&B\leq B_0\nonumber\\
&&M\in\mathbf{R}^{+}\nonumber\\
&&B\in \mathbf{R}^{+},\nonumber
\end{eqnarray}
where $\mathbf{R}^{+}$ is the collection of the positive real value.
$J_2$ is a nonlinear programming (NLP) problem with linear
constraints. At present, a commonly used method to solve such a
problem is sequential linear programming (SLP). SLP consists of
linearizing the objective in a region around a nominal operating
point by a Taylor series expansion. The resulting linear programming
problem is then solved by standard methods such as the interior
point method. Some softwares, i.e. \emph{Lingo} and \emph{Matlab},
usually solve the NLP problem by SLP, so we can use them to resolve
$J_2$ directly. Assuming $(M^{\dag},B^{\dag})$ is the solution to
$J_2$. Let $M_1=\lceil M^{\dag}\rceil$ and $M_2=\lfloor
M^{\dag}\rfloor$ be the candidate transmission modes, and
$B_1=\lceil B^{\dag}\rceil$ and $B_2=\lfloor B^{\dag}\rfloor$ be the
candidate feedback amounts. Then, select the combination that has
the maximum feedback utility function from $(M_1,B_1)$, $(M_1,B_2)$,
$(M_2,B_1)$ and $(M_2,B_2)$ as the solution to the original
optimization problem. Although the solution of this algorithm may
not be optimal, it can achieve a tradeoff between the performance
and computation complexity. With respect to the optimal numerical
searching algorithm mentioned above, which needs to compare
$(N_t-2)(B_0+1)$ feedback utility functions, the proposed algorithm
has a quite low complexity, especially when $N_t$ and $B_0$ are
large. It is worth pointing out that this algorithm is performed
only when channel condition $\alpha$ varies, but not during each
time slot, so it can be run offline.

\emph{Remark1}: In this paper, we consider the case with one primary
user. Actually, the proposed scheme can generalize to the case of
multiple primary users directly. The impact of multiple primary
users lies in three folds. First, assuming there are $L$ primary
users, the maximum transmission mode is $N_t-L$, because $L$ spatial
degrees of freedom is used to mitigate the interferences to $L$
primary users. In other words, if $L\geq N_t$, no cognitive user is
allowed to access the licensed spectrum. Second, transmit power is
constrained by $L$ interference channels simultaneously. For
simplicity, assuming cognitive BS purchases the same feedback amount
$B$ from all $L$ primary users, transmit power is upper bounded by
the interference channel that has the largest path loss $\alpha$.
Third, cognitive BS needs to purchase feedback from $L$ primary
users. The feedback cost increases linearly with $L$ and the average
sum rate decreases due to the decrease of spatial degree of freedom,
so the maximum feedback utility function may become negative. Under
this condition, cognitive network may be unwilling to use this
spectrum.

\emph{Remark2}: The proposed scheme also can be generalized to the
case of variable $N_t$ by taking $N_t$ as an optimization variable,
so that the performance of cognitive network may be further
improved. Under this condition, the focused problem in this paper
becomes a subproblem of the whole optimization problem. since $N_t$
is an integer variable, we can first derive all the combinations of
$M$ and $B$ when scaling $N_t$ from $2$ to the number of antennas at
cognitive BS minus $L$ by the proposed scheme. Then, find the
optimal combination $(M, B)$ with the maximum feedback utility
function, the corresponding $N_t$ is the optimal antenna number.

Due to relatively long distance between primary receiver and
cognitive BS, there may be more or less feedback delay during
quantized CSI conveyance, resulting in CSI mismatch when designing
ZFBF, and thus interference constraint may be unsatisfied if
outdated CSI is used directly. Herein, we give an investigation of
the impact of feedback delay on feedback utility function and then
present a robust mode selection scheme based on limited feedback.
Following \cite{ImperfectCSI}, we model the CSI as
\begin{equation}
\textbf{g}_r=\sqrt{\rho}\textbf{g}_o+\sqrt{1-\rho}\textbf{e},\label{eqn21}
\end{equation}
where $\textbf{g}_r$ and $\textbf{g}_o$ are the real and the
outdated CSI, respectively. $\textbf{e}$ is the error vector due to
feedback delay with i.i.d. zero mean and unit variance complex
Gaussian entries, and $\rho$ is the correlation coefficient, which
is dependent of normalized frequency shift and delay duration.
Clearly, the larger the correlation coefficient $\rho$, the less the
CSI mismatch. Substituting (\ref{eqn21}) into (\ref{eqn4}), the
interference term is transformed as
\begin{eqnarray}
I&=&P/M\alpha\sum\limits_{m=1}^M|\textbf{g}_r^H\textbf{w}_{K_m}|^2\nonumber\\
&=&P/M\alpha\sum\limits_{m=1}^M|\sqrt{\rho}\textbf{g}_o^H\textbf{w}_{K_m}+\sqrt{1-\rho}\textbf{e}^H\textbf{w}_{K_m}|^2\nonumber\\
&=&P/M\alpha\sum\limits_{m=1}^M|\sqrt{a\rho}\textbf{s}^H\textbf{w}_{K_m}+\sqrt{1-\rho}\textbf{e}^H\textbf{w}_{K_m}|^2.\label{eqn22}
\end{eqnarray}
Under this condition, the average AIC can be computed as
\begin{eqnarray}
\bar{I}&=&P\alpha
E\left[|\sqrt{a\rho}\textbf{s}^H\textbf{w}_{K_m}+\sqrt{1-\rho}\textbf{e}^H\textbf{w}_{K_m}|^2\right]\nonumber\\
&=&P\alpha\left(E\left[a\rho|\textbf{s}^H\textbf{w}_{K_m}|^2\right]+E\left[(1-\rho)|\textbf{e}^H\textbf{w}_{K_m}|^2\right]\right)\label{eqn23}\\
&<&P\alpha\left(\frac{N_t\rho}{N_t-1}2^{-\frac{B}{N_t-1}}+(1-\rho)\right)\label{eqn24}\\
&=&\frac{\alpha
PN_t}{N_t-1}2^{-\frac{B}{N_t-1}}\left(1+\left(1-\rho\right)\left(\frac{N_t-1}{N_t}2^{\frac{B}{N_t-1}}-1\right)\right),\nonumber\\\label{eqn25}
\end{eqnarray}
where (\ref{eqn23}) holds true because $\textbf{s}$ and $\textbf{e}$
are independent of each other, (\ref{eqn24}) following from the fact
that $|\textbf{e}^H\textbf{w}_{K_m}|^2$ is $\chi_2^2$ distributed
and $E\left[|\textbf{e}^H\textbf{w}_{K_m}|^2\right]=1$. Notice that
the average interference without CSI mismatch is $\frac{\alpha
PN_t}{N_t-1}2^{-\frac{B}{N_t-1}}$, if
$\frac{N_t-1}{N_t}2^{\frac{B}{N_t-1}}>1$, the CSI mismatch leads to
the increase of average interference. In other words, with the same
average interference constraint and feedback amount, the available
maximum transmit power decreases accordingly due to CSI mismatch.
Otherwise, if $(N_t-1)/N_t 2^{B/(N_t-1)}<1$, the CSI mismatch will
decrease the average interference, so that cognitive BS can use
higher transmit power to improve the performance. Hence, the upper
bound of transmit power in presence of delay while satisfying AIC
can be expressed as
\begin{equation}
P=\frac{I_{\textrm{AIC}}}{\alpha\left(\frac{N_t\rho}{N_t-1}2^{-\frac{B}{N_t-1}}+(1-\rho)\right)}.\label{eqn26}
\end{equation}
Substituting (\ref{eqn26}) into (\ref{eqn9}), we can derive the
feedback utility function in presence of feedback delay, and thus
the corresponding optimal $B$ and $M$ can be obtained by using the
above proposed algorithm.

\section{Asymptotic Analysis}
In this section, for the sake of deriving the optimal $(M,B)$ and
evaluating the performance easily, we analyze the asymptotic
characteristics of feedback utility function in some extreme cases.

\subsection{High AIC and/or Small $\alpha$}
If primary receiver can bear sufficiently high AIC and/or $\alpha$
is sufficiently small due to large distance between cognitive BS and
primary receiver, such that AIC can still be met even with $B=0$ and
arbitrarily high transmit power. Under such a condition, it is clear
that $B=0$ is optimal in the sense of maximizing feedback utility
function. In other words, without the help from primary receiver,
cognitive network can work freely. It is equivalent to the
traditional mode selection on broadcast channel. Assuming a certain
transmit power $P^{'}$ is used at cognitive BS, the optimal
transmission mode is determined based on the average sum rate of
cognitive users
\begin{equation}
R(M)=\frac{M\log_2(e)}{\delta^{M-1}}I_3\left(\sigma^2M/P^{'},
\delta^{-1}, M-1\right).\label{eqn17}
\end{equation}
Due to the complex expression, numerical search method is also used
to derive the optimal $M$. However, because of only one searching
variable, the computation complexity can be reduced significantly.

\subsection{Low AIC and/or Large $\alpha$}
In the case that primary receiver has a quite strict constraint on
AIC and/or $\alpha$ is sufficiently large, the second term of the
denominator in (\ref{eqn12}) can be neglected compared to the first
one. Hence, $SINR_{K_m}$ is reduced to
\begin{equation}
SINR_{K_m}=\frac{\|\textbf{h}_{K_m}\|^2\beta(1,N_t-1)}{\eta
2^{-\frac{B}{N_t-1}}},\label{eqn18}
\end{equation}
and the corresponding pdf is
\begin{equation}
p(x)=\eta M2^{-\frac{B}{N_t-1}}\exp\left(-\eta
M2^{-\frac{B}{N_t-1}}x\right).\label{eqn19}
\end{equation}
Therefore, the feedback utility function under this condition can be
computed as
\begin{eqnarray}
f(M,B)&=&M\int_0^\infty\log_2\left(1+x\right)p(x)dx-\mu
B\nonumber\\
&=&M\log_2(e)\exp\left(\eta
M2^{-\frac{B}{N_t-1}}\right)E_1\left(\eta
M2^{-\frac{B}{N_t-1}}\right)\nonumber\\
&-&\mu B.\label{eqn20}
\end{eqnarray}
It is known from (\ref{eqn20}) that, $f(M,B)$ is a monotone
increasing function of $M$, so that $M=N_t-1$ is optimal. On the
other hand, $f(M,B)$ is a monotone decreasing function of $B$, then
$B=0$ is optimal. In fact, this case is equivalent to a
noise-limited system, the help from primary receiver makes no sense
and full multiplexing $(M=N_t-1)$ is propitious to improve the sum
rate, so that it is reasonable to choose $B=0$ and $M=N_t-1$.

\subsection{Large C}
When $C$ is quite large, but still fixed,
$b=\sin^2(\angle(\tilde{\textbf{h}}_{K_m},\check{\textbf{h}}_{K_m}))\approx0$.
In other words, the inter-user interference can be neglected. Under
this condition, $SINR_{K_m}$ has the same expression as
(\ref{eqn18}), so that $M=N_t-1$ and $B=0$ are optimal.

\section{Numerical Results}
To examine the effectiveness of the proposed feedback efficient
adaptive mode selection scheme in limited cooperative multiuser
cognitive networks, we present several numerical results in
different scenarios. For all scenarios, we set $N_t=5$ considering
the maximum number of BS antennas in LTE-A is 8, $B_0=4$ due to the
complexity constraint on codeword selection at primary receiver, and
$\sigma^2=1$. Additionally, we use FUF$_{\textrm{max}}$ to denote
the maximum feedback utility function when selecting the optimal
transmission mode $M_{\textrm{opt}}$ and feedback bits
$B_{\textrm{opt}}$, and use OA and SA to denote the optimal and
suboptimal algorithms, respectively. Interestingly, it is found that
the suboptimal algorithm has the same optimization results as the
optimal one in all scenarios, so that we could replace the optimal
algorithm with the suboptimal one in practical networks.

\begin{figure}[h] \centering
\includegraphics [width=0.5\textwidth] {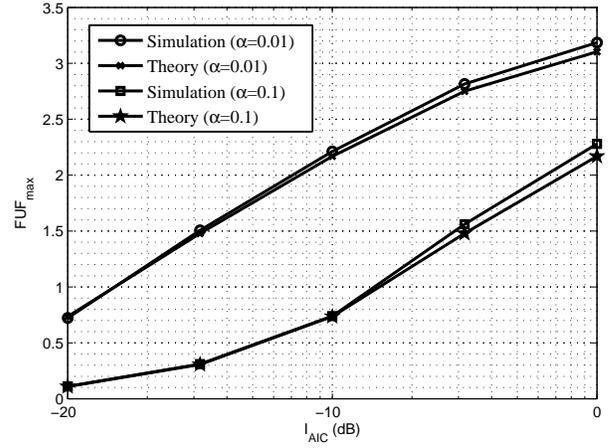}
\caption {Comparison of maximum FUF of feedback efficient adaptive
mode selection based on numerical simulation and theoretical
deduction.} \label{Fig2}
\end{figure}

First, we testify the validity of the theoretical claims through
numerical simulation with different path losses. In this case, we
fix $C=2$ and $\mu=0.1$ for convenience. As seen in Fig.\ref{Fig2},
in the whole region of interference constraint, the theoretical and
simulation results are quite identical. Especially when
$I_{\textrm{AIC}}<-10$dB, the two ones are nearly coincident for
different path losses. Meanwhile, it is found that path loss
$\alpha$ has a great impact on mode selection and feedback bits, and
thus maximum feedback utility function. As a simple example, at
$I_{\textrm{AIC}}=-10$dB, FUF$_{\textrm{max}}$ for $\alpha=0.01$ has
a performance gain of about $1.5$dB over that for $\alpha=0.1$. This
is because smaller $\alpha$ leads to higher transmit power at
cognitive BS for a given interference constraint. In addition, as
shown in Tab.\ref{Tab1}, in the case of high AIC and small $\alpha$,
$B=0$ is optimal, while $B=0$ and $M=N_t-1=4$ can maximize feedback
utility function for the case of low AIC and large $\alpha$, which
are consistent with the asymptotic analysis results.

\begin{table}\centering
\caption{The optimal transmission mode and feedback bits with
different path losses} \label{Tab1}
\begin{tabular}{|c|c|c|c|c|c|c|c|}\hline
& & $I_{\textrm{AIC}}$ & -20dB & -15dB & -10dB & -5dB & 0dB \\
\hline\hline & & $B_{\textrm{opt}}$ & 0 & 4 & 1 & 0 & 0\\
\cline{3-8} $\alpha=0.01$ & $\textmd{OA}$ &$M_{\textrm{opt}}$ & 4 & 2 & 2 & 2 & 2\\
\cline{2-8} & & $B_{\textrm{opt}}$ & 0 & 4 & 1 & 0 & 0\\
\cline{3-8}  & $\textmd{SA}$ &$M_{\textrm{opt}}$ & 4 & 2 & 2 & 2 & 2\\
\hline\hline & & $B_{\textrm{opt}}$ & 0 & 0 & 0 & 4 & 1\\
\cline{3-8} $\alpha=0.1$ & $\textmd{OA}$ & $M_{\textrm{opt}}$ & 4 & 4 & 4 & 2 & 2\\
\cline{2-8} & & $B_{\textrm{opt}}$ & 0 & 0 & 0 & 4 & 1\\
\cline{3-8}  & $\textmd{SA}$ &
$M_{\textrm{opt}}$ & 4 & 4 & 4 & 2 & 2\\
\hline
\end{tabular}
\end{table}

\begin{figure}[h] \centering
\includegraphics [width=0.5\textwidth] {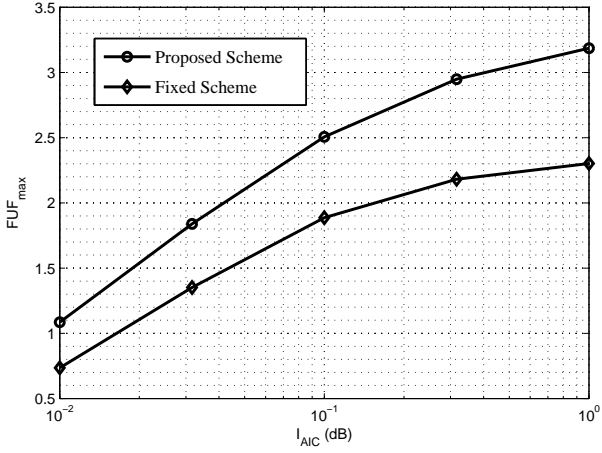}
\caption {Comparison of maximum FUF of proposed scheme and fixed
scheme.} \label{Fig6}
\end{figure}

Secondly, we show the performance gain from cooperative feedback by
comparing the performance of the proposed scheme and a fixed scheme
with $\alpha=0.01$ and $C=2$. As the name implies, the fixed scheme
sets $M=N_t=5$ and $B=0$, namely without cooperation from primary
network. Note that the fixed scheme has a larger multiplexing gain,
since there is no need of extra spatial degree of freedom for
mitigating the interference to primary receiver. In spite of this,
the proposed scheme performs better than the fixed scheme obviously
in the whole range of $I_{\textrm{AIC}}$. As seen in Fig.\ref{Fig6},
as $I_{\textrm{AIC}}$ relaxes, the performance gain becomes larger
and larger. It is worth pointing out that from the perspective of
average transmission rate, the proposed scheme has more gains, since
FUF$_{\textrm{max}}$ takes into consideration of feedback cost.
Therefore, the proposed scheme can effectively improve the
performance.

\begin{figure}[h] \centering
\includegraphics [width=0.5\textwidth] {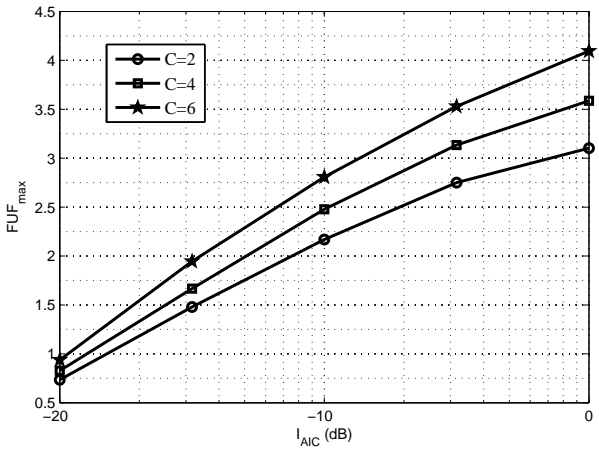}
\caption {Comparison of maximum FUF of feedback efficient adaptive
mode selection with different feedback amount.} \label{Fig3}
\end{figure}

Next, we investigate the impact of feedback amount of cognitive user
on the optimal mode selection, feedback bits and
FUF$_{\textrm{max}}$. In this case, we fix $\alpha=0.01$ and
$\mu=0.1$. Intuitively, the larger $C$ is conducive to improve
FUF$_{\textrm{max}}$, since more accurate CSI can further decrease
the inter-user interference. As seen in Fig.\ref{Fig3}, at
$I_{\textrm{AIC}}=-10$dB, FUF$_{\textrm{max}}$ for $C=6$ has a
performance gain of about $1.0$dB over that for $C=2$, and the gain
increases with a lower AIC requirement. In addition, as analyzed in
Section IV, the impacts of a small $\alpha$ and a large $C$ are
contrary, so that $B_{\textrm{opt}}$ is unequal to 0 when $C=6$ and
$I_{\textrm{AIC}}\leq-10$dB, as shown in Tab.\ref{Tab2}. However, as
$I_{\textrm{AIC}}$ increases, $B_{\textrm{opt}}$ becomes small
gradually, and $B_{\textrm{opt}}=0$ when $I_{\textrm{AIC}}=0$dB.

\begin{table}\centering
\caption{The optimal transmission mode and feedback bits with
different feedback amount} \label{Tab2}
\begin{tabular}{|c|c|c|c|c|c|c|c|}\hline
& & $I_{\textrm{AIC}}$ & -20dB & -15dB & -10dB & -5dB & 0dB \\
\hline\hline & & $B_{\textrm{opt}}$ & 0 & 4 & 1 & 0 & 0
\\\cline{3-8} $C=2$& $\textmd{OA}$ & $M_{\textrm{opt}}$ & 4 & 2 & 2 & 2 & 2\\
\cline{2-8} & & $B_{\textrm{opt}}$ & 0 & 4 & 1 & 0 & 0
\\\cline{3-8} & $\textmd{SA}$ & $M_{\textrm{opt}}$ & 4 & 2 & 2 & 2 & 2\\
\hline\hline & & $B_{\textrm{opt}}$ & 4 & 4 & 4 & 0 & 0
\\\cline{3-8} $C=4$ & $\textmd{OA}$ & $M_{\textrm{opt}}$ & 4 & 2 & 2 & 2 &2\\
\cline{2-8} & & $B_{\textrm{opt}}$ & 4 & 4 & 4 & 0 & 0
\\\cline{3-8} & $\textmd{SA}$ & $M_{\textrm{opt}}$ & 4 & 2 & 2 & 2 &2\\
\hline\hline
 & & $B_{\textrm{opt}}$ & 4 & 4 & 4 & 1 & 0
\\\cline{3-8} $C=6$ & $\textmd{OA}$ & $M_{\textrm{opt}}$ & 4 & 4 & 2 & 2 & 2\\
\cline{2-8} & & $B_{\textrm{opt}}$ & 4 & 4 & 4 & 1 & 0
\\\cline{3-8} & $\textmd{SA}$ & $M_{\textrm{opt}}$ & 4 & 4 & 2 & 2 & 2\\
\hline
\end{tabular}
\end{table}

\begin{figure}[h] \centering
\includegraphics [width=0.5\textwidth] {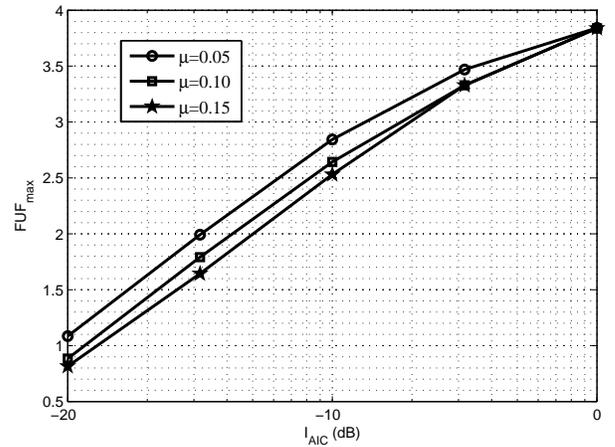}
\caption {Comparison of maximum FUF of feedback efficient adaptive
mode selection with different pricing factors.} \label{Fig4}
\end{figure}

Then, we compares the FUF$_{\textrm{max}}$s of feedback efficient
adaptive mode selection with different pricing factors. In this
scenario, we fix $C=5$ and $\alpha=0.01$. As seen in Fig.\ref{Fig4},
$\mu$ has a relatively slight impact on the FUF$_{\textrm{max}}$.
Especially when $I_{\textrm{AIC}}$ becomes less loose, such as
$I_{\textrm{AIC}}>-5$dB, FUF$_{\textrm{max}}$ is hardly changed as
the pricing factor $\mu$ increases. This is because cognitive BS
purchases less feedback bits and transmission mode keeps constant.
Once pricing factor $\mu$ rises, as shown in Tab.\ref{Tab3}, both
available transmit power and feedback cost descend simultaneously,
and thus FUF$_{\textrm{max}}$ is not so unsensitive to $\mu$.
Additionally, for a given $\mu$, with the increase of AIC, the
required $B_{\textrm{opt}}$ decreases gradually, and reduces to 0
finally when $I_{\textrm{AIC}}$ is large enough, and this is
consistent of the asymptotic analysis result. With the increase of
$\mu$, $B_{\textrm{opt}}$ is equal to 0 even with a strict AIC.
Meanwhile, the strict AIC leads to full multiplexing, namely
$M_{\textrm{opt}}=N_t-1$, because the cognitive network is
interference-limited under such a condition. These conclusions are
helpful to determine the extent of cooperation in practical
networks.

\begin{table}\centering
\caption{The optimal transmission mode and feedback bits with
different pricing factors} \label{Tab3}
\begin{tabular}{|c|c|c|c|c|c|c|c|}\hline
& & $I_{\textrm{AIC}}$ & -20dB & -15dB & -10dB & -5dB & 0dB \\
\hline\hline & & $B_{\textrm{opt}}$ & 4 & 4 & 4 & 4 & 1
\\\cline{3-8} $\mu=0.05$ & $\textmd{OA}$ & $M_{\textrm{opt}}$ & 4 & 4 & 2 & 2 & 2\\
\cline{2-8} & & $B_{\textrm{opt}}$ & 4 & 4 & 4 & 4 & 1
\\\cline{3-8}  & $\textmd{SA}$ & $M_{\textrm{opt}}$ & 4 & 4 & 2 & 2 & 2\\
\hline\hline
 & & $B_{\textrm{opt}}$ & 4 & 4 & 4 & 0 & 0
\\\cline{3-8} $\mu=0.10$ & $\textmd{OA}$ & $M_{\textrm{opt}}$ & 4 & 4 & 2 & 2 & 2\\
\cline{2-8} & & $B_{\textrm{opt}}$ & 4 & 4 & 4 & 0 & 0
\\\cline{3-8} & $\textmd{SA}$ & $M_{\textrm{opt}}$ & 4 & 4 & 2 & 2 & 2\\
\hline\hline
 & & $B_{\textrm{opt}}$ & 0 & 0 & 0 & 0 & 0
\\\cline{3-8} $\mu=0.15$ & $\textmd{OA}$ & $M_{\textrm{opt}}$ & 4 & 4 & 2 & 2 & 2\\
\cline{2-8} & & $B_{\textrm{opt}}$ & 0 & 0 & 0 & 0 & 0
\\\cline{3-8} & $\textmd{SA}$ & $M_{\textrm{opt}}$ & 4 & 4 & 2 & 2 & 2\\
\hline
\end{tabular}
\end{table}

\begin{figure}[h] \centering
\includegraphics [width=0.5\textwidth] {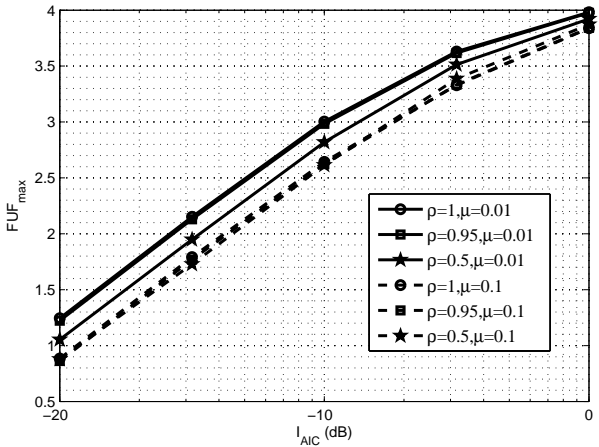}
\caption {Comparison of maximum FUF of feedback efficient adaptive
mode selection with different correlation coefficients.}
\label{Fig5}
\end{figure}

Finally, we consider the scenario in presence of feedback delay
during conveying interference CSI, and the delay for feedback link
of cognitive network can be neglected due to the relatively short
access distance. Similarly, we fix $C=5$ and $\alpha=0.01$ in this
scenario. Surprisingly, it is seen from Fig.\ref{Fig5} that
FUF$_{\textrm{max}}$ is quite unsensitive to feedback delay. For
$\mu=0.1$, even with large feedback delay, such as $\rho=0.5$,
FUF$_{\textrm{max}}$ coincides with that of ideal case, namely
$\rho=1$. Furthermore, FUF$_{\textrm{max}}$ for $\rho=0.5$ may be
slightly better than that of $\rho=1$ under some conditions, such as
a loose AIC, which reconfirms to the theoretical claim earlier.
Meanwhile, Tab.\ref{Tab4} shows that feedback delay will not change
the optimal transmission mode when $\mu=0.1$. If $\mu$ is sufficient
small, such as $\mu=0.01$, as many feedback bits as possible is used
due to such a low price. Under this condition, the case of
$\rho=0.5$ has a performance loss, because it can not compensate the
loss caused by decreasing feedback bits. However, in practical
networks, cognitive networks is impossible to obtain the cooperation
on such a low cost, so the proposed feedback efficient adaptive mode
selection is robust under various scenarios.

\begin{table}\centering
\caption{The optimal transmission mode and feedback bits with
different correlation coefficients} \label{Tab4}
\begin{tabular}{|c|c|c|c|c|c|c|c|}\hline
& & $I_{\textrm{AIC}}$ & -20dB & -15dB & -10dB & -5dB & 0dB \\
\hline\hline & & $B_{\textrm{opt}}$ & 4 & 4 & 4 & 4 & 4
\\\cline{3-8} $\rho=1$ & $\textmd{OA}$ & $M_{\textrm{opt}}$ & 4 & 4 & 2 & 2 & 2\\
\cline{2-8} $\mu=0.01$& & $B_{\textrm{opt}}$ & 4 & 4 & 4 & 4 & 4
\\\cline{3-8} & $\textmd{SA}$ & $M_{\textrm{opt}}$ & 4 & 4 & 2 & 2 & 2\\
\hline\hline & & $B_{\textrm{opt}}$ & 4 & 4 & 4 & 4 & 4
\\\cline{3-8} $\rho=0.95$  & $\textmd{OA}$ & $M_{\textrm{opt}}$ & 4 & 4 & 2 & 2 & 2\\
\cline{2-8} $\mu=0.01$ & & $B_{\textrm{opt}}$ & 4 & 4 & 4 & 4 & 4
\\\cline{3-8}  & $\textmd{SA}$ & $M_{\textrm{opt}}$ & 4 & 4 & 2 & 2 & 2\\
\hline\hline & & $B_{\textrm{opt}}$ & 4 & 4 & 4 & 4 & 4
\\\cline{3-8}  $\rho=0.5$ & $\textmd{OA}$ & $M_{\textrm{opt}}$ & 4 & 4 & 2 & 2 & 2\\
\cline{2-8} $\mu=0.01$ & & $B_{\textrm{opt}}$ & 4 & 4 & 4 & 4 & 4
\\\cline{3-8}   & $\textmd{SA}$ & $M_{\textrm{opt}}$ & 4 & 4 & 2 & 2 & 2\\
\hline\hline & & $B_{\textrm{opt}}$ & 4 & 4 & 4 & 0 & 0
\\\cline{3-8} $\rho=1$  & $\textmd{OA}$ & $M_{\textrm{opt}}$ & 4 & 4 & 2 & 2 & 2\\
\cline{2-8} $\mu=0.1$ & & $B_{\textrm{opt}}$ & 4 & 4 & 4 & 0 & 0
\\\cline{3-8} & $\textmd{SA}$ & $M_{\textrm{opt}}$ & 4 & 4 & 2 & 2 & 2\\
\hline\hline
 & & $B_{\textrm{opt}}$ & 4 & 4 & 4 & 0 & 0
\\\cline{3-8} $\rho=0.95$  & $\textmd{OA}$ & $M_{\textrm{opt}}$ & 4 & 4 & 2 & 2 & 2\\
\cline{2-8} $\mu=0.1$ & & $B_{\textrm{opt}}$ & 4 & 4 & 4 & 0 & 0
\\\cline{3-8} & $\textmd{SA}$ & $M_{\textrm{opt}}$ & 4 & 4 & 2 & 2 & 2\\
\hline\hline
 & & $B_{\textrm{opt}}$ & 0 & 0 & 0 & 0 & 0
\\\cline{3-8} $\rho=0.5$ & $\textmd{OA}$ & $M_{\textrm{opt}}$ & 4 & 4 & 2 & 2 & 2\\
\cline{2-8} $\mu=0.1$ & & $B_{\textrm{opt}}$ & 0 & 0 & 0 & 0 & 0
\\\cline{3-8} & $\textmd{SA}$ & $M_{\textrm{opt}}$ & 4 & 4 & 2 & 2 & 2\\
\hline
\end{tabular}
\end{table}

\section{Conclusion}
A major contribution of this paper is to propose a limited
cooperation between primary and cognitive networks, so as to take
advantage of the benefit of multiantenna at cognitive BS through
ZFBF. In order to utilize the feedback information from primary
receiver effectively, this paper further proposes to maximize the
feedback utility function while satisfying the interference
constraint by optimizing transmission mode and feedback bits. With
the goal of enhancing the usability of the present adaptive mode
selection scheme in practical networks, we investigate the effect of
feedback delay on feedback utility function and derive the
corresponding optimal transmission scheme. In addition, we present
some low-complexity schemes through asymptotic characteristics
analysis.


\begin{thebibliography}{1}
\bibitem{FCC}
FCC, ``Spectrum policy taks force report," ET Docket 02-155, Nov.
2002.

\bibitem{Mitola}
I. J. Mitola, ``Software radio: survay, critical evaluation and
future direction," \emph{IEEE  Aerosp. Elecron. Syst. Mag.}, vol. 8,
pp. 25-31, Apr. 1993.

\bibitem{Haykin}
B. Wang, and K. J. R. Liu, ``Advances in cognitive radio networks: a
survey," \emph{IEEE J. Sel. Topics Signal Process.}, vol. 5, no. 1,
pp. 5-23, Feb. 2011.

\bibitem{sensing1}
S. Haykin, D. J. Thomson, and J. H. Reed, ``Spectrum sensing for
cognitive radio," \emph{Proceedings of the IEEE}, vol. 97, no. 5,
pp. 849-877, May 2009.

\bibitem{sensing2}
S. Makeki, and G. Leus, ``Censored truncated sequential spectrum
sensing cog cognitive radio networks," \emph{IEEE J. Sel. Areas
Commun.}, vol. 31, no. 3, pp. 364-378, Mar. 2013.

\bibitem{sensing3}
R. Deng, J. Chen, C. Yuen, P. Cheng, Y. Sun, ``Energy-efficient
cooperative spectrum sensing by optimal scheduling in sensor-aided
cognitive radio networks," \emph{IEEE Trans. Veh. Tech.}, vol. 61,
no. 2, pp. 716-725, Feb. 2012.

\bibitem{PA}
Y. Chen, G. Yu, Z. Zhang, H. Chen, and P. Qiu, ``On cognitive radio
networks with opportunistic power control strategies in fading
channels," \emph{IEEE Trans. Wireless Commun.}, vol. 7, no. 7, pp.
2752-2761, Jul. 2008.

\bibitem{Multiantenna1}
V. D. Papoutsis and A. P. Stamouli, ``Chunk-based resource
allocation in multicast MISO-OFDMA with average BER constraint,"
\emph{IEEE Commun. Lett.}, vol. 17, no. 2, pp. 317-320, Feb. 2013.

\bibitem{Multiantenna2}
V. D. Papoutsis and S. A. Kotsopoulos, ``Chunk-based resource
allocation in distributed MISO-OFDMA systems with fairness
guarantee," \emph{IEEE Commun. Lett.}, vol. 15, no. 4, pp. 377-379,
Apr. 2011.

\bibitem{CogMIMO1}
F. Gao, R. Zhang, Y-C. Liang, X. Wang, ``Design and learning-based
MIMO cognitive radio systems," \emph{IEEE Trans. Veh. Tech.}, vol.
59, no. 4, pp. 1707-1720, May 2010.

\bibitem{CogMIMO4}
X. Chen, and H-H. Chen, ``Interference-aware resource control in
multi-antenna cognitive ad hoc networks with heterogeneous delay
constraints," \emph{IEEE Commun. Lett.}, vol. 17, no. 6, pp.
1184-1187, Apr. 2013.

\bibitem{CogMIMO2}
X. Zeng, Q. Li, Q. Zhang, and J. Qin, ``Joint beamforming and
antenna subarray formation for MIMO cognitive radio," \emph{IEEE
Signal Process. Lett.}, vol. 20, no. 5, pp. 479-482, May 2013.

\bibitem{CogMIMO3}
J. Noh, and S. Oh, ``Beamforming in a multi-user cognitive radio
system with partial channel state information," \emph{IEEE Trans.
Wireless Commun.}, vol. 12, no. 2, pp. 616-625, Feb. 2013.

\bibitem{MA1}
R. Zhang, and Y-C. Liang, ``Exploiting multi-antennas for
opportunistic spectrum sharing in cognitive radio networks,"
\emph{IEEE J. Sel. Topics Signal Process.}, vol. 2, no. 1, pp.
88-102, Feb. 2008.

\bibitem{MA2}
X. Chen, and C. Yuen, ``Efficient resource allocation in rateless
coded MU-MIMO cognitive radio network with QoS provisioning and
limited feedback," \emph{IEEE Trans. Veh. Tech.}, vol. 62, no. 1,
pp. 395-399, Jan. 2013.

\bibitem{MA3}
L. Zhang, Y-C. Liang, Y. Xin, and H. Poor, ``Robust cognitive
beamforming with partial channel state information," \emph{IEEE
Trans. Wireless Commun.}, vol. 8, no. 8, pp. 4143-4152, Aug. 2009.

\bibitem{Multiuser1}
K. Hamdi, W. Zhang, K. B. Letaief, ``Joint beamforming and
scheduling in cognitive radio network," in \emph{Proc. IEEE
Globecom}, pp. 2977-2981, Nov. 2007.

\bibitem{Multiuser2}
R. Zhang, and Y-C. Liang, ``Investigation on multiuser diversity in
spectrum sharing based cognitive radio networks," \emph{IEEE Commun.
Lett.}, vol. 14, no. 2, pp. 133-135, Feb. 2010.

\bibitem{Multiuser3}
M. H. Islam, Y-C. Liang, and A. T. Hoang, ``Joint power control and
beamforming for cognitive radio networks," \emph{IEEE Trans.
Wireless Commun.}, vol. 7, no. 7, pp. 2415-249, Jul. 2008.

\bibitem{Feedback1}
D. J. Love, R. W. Heath Jr., V. K. N. Lau, D. Gesbert, B. D. Rao,
and M. Andrews, ``An overview of limited feedback in wireless
communication systems," \emph{IEEE J. Sel. Areas Commun.}, vol. 26,
no. 8, pp. 1341-1365, Oct. 2008.

\bibitem{Feedback2}
H. Yang, J. Chun, and Y. Choi, ``Codebook-based
lattice-reduction-aided precoding for limited-feedback coded MIMO
systems," \emph{IEEE Trans. Commun.}, vol. 60, no. 2, pp. 510-524,
Feb. 2012.

\bibitem{CooperativeFeedback1}
K. Huang, and R. Zhang, ``Cooperative feedback for multiantenna
cognitive radio networks," \emph{IEEE Trans. Signal Process.}, vol.
59, no. 2, pp. 747-758, Feb. 2011.

\bibitem{CooperativeFeedback2}
S. Ganesan, M. Sellathurai, and T. Ratnarajah, ``Opportunistic
interference projection in cognitive MIMO radio with multiuser
diversity," \emph{IEEE New Frontiers in Dynamic Spectrum}, pp. 1-6,
Apr. 2010.

\bibitem{ModeSelection1}
L. Liu, R. Chen, S. Geirhofer, K. Sayana, Z. Shi, and Y. Zhou,
``Downlink MIMO in LTE-advanced: SU-MIMO vs. MU-MIMO," \emph{IEEE
Commun. Mag.}, vol. 50, no. 2, pp. 140-147, Feb. 2012.

\bibitem{ModeSelection2}
X. Chen, Z. Zhang, S. Chen, and C. Wang, ``Adaptive mode selection
for multiuser MIMO downlink employing rateless codes with QoS
provioning," \emph{IEEE Trans. Wireless Commun.}, vol. 11, no. 2,
pp. 790-799, Feb. 2012.

\bibitem{Modeselection3}
J. Zhang, M. Kountouris, J. G. Andrews, and R. W. Heath Jr.,
``Multi-mode transmission for the MIMO broadcast channel with
imperfect channel state information," \emph{IEEE Trans. Commun.},
vol. 59, no. 3, pp. 803-814, Mar. 2011.

\bibitem{AIC1}
R. Zhang, ``On peak versus average intererence power constraints for
protecting primary users in cognitive radio networks," \emph{IEEE
Trans. Wireless Commun.}, vol. 8, no. 4, pp. 2112-2120, Apr. 2009.

\bibitem{AIC2}
C-X. Wang, X. Hong, H-H. Chen, and J. Thompson, ``On capacity of
cognitive radio networks with average interference power
constraints," \emph{IEEE Trans. Wireless Commun.}, vol. 8, no. 4,
pp. 1620-1625, Apr. 2009.

\bibitem{Stackelberg}
X. Kang, R. Zhang, and M. Motani, ``Price-based resource allocation
for spectrum-sharing femtocell networks: a stackelberg game,"
\emph{IEEE J. Sel. Areas Commun.}, vol. 30, no. 3, pp. 538-549, Apr.
2012.

\bibitem{RVQ}
N. Jindal, ``MIMO Broadcast channels with finite-rate feedback,"
\emph{IEEE Trans. Inf. Theory}, vol. 52, no. 11, pp. 5045-5060, Nov.
2006.

\bibitem{QCA}
K. K. Mukkavilli, A. Sabharwal, E. Erkip, and B. Aazhang, ``On
beamforming with finite rate feedback in multiple-antenna systems,"
\emph{IEEE Trans. Inf. Theory}, vol. 49, no. 10, pp. 2562-2579, Oct.
2003.

\bibitem{ImperfectCSI}
Y. Isukapalli, and B. D. Rao, ``Finite rate feedback for spatially
and temporally correlated MISO channels in the presence of
estimation errors and feedback delay," in \emph{Proc. IEEE
GLOBECOM}, pp. 2791-2795, Nov. 2007.

\end{thebibliography}
\end{document}